# Molecular crowding in single eukaryotic cells: using cell environment biosensing and single-molecule optical microscopy to probe dependence on extracellular ionic strength, local glucose conditions, and sensor copy number

Jack W Shepherd[1,a,b], Sarah Lecinski[1,a], Jasmine Wragg[c], Sviatlana Shashkova[a,d], Chris MacDonald[b], Mark C Leake[a,b]

[1]These authors contributed equally

[a] Department of Physics, University of York, York, YO10 5DD

[b] Department of Biology, University of York, York, YO10 5DD

[c] School of Natural Sciences, University of York, York, YO10 5DD

[d] Department of Microbiology and Immunology, Institute for Biomedicine, Sahlgrenska Academy, University of Gothenburg, 405 30 Gothenburg, Sweden

For correspondence: Jack W Shepherd, Department of Physics, University of York, York, YO10 5DD. E-mail: jack.shepherd@york.ac.uk

**Abstract**

The physical and chemical environment inside cells is of fundamental importance to all life but has traditionally been difficult to determine on a subcellular basis. Here we combine cutting-edge genomically integrated FRET biosensing to readout localized molecular crowding in single live yeast cells. Confocal microscopy allows us to build subcellular crowding heatmaps using ratiometric FRET, while whole-cell analysis demonstrates crowding is reduced when yeast is grown in elevated glucose concentrations. Simulations indicate that the cell membrane is largely inaccessible to these sensors and that cytosolic crowding is broadly uniform across each cell over a timescale of seconds. Millisecond single-molecule optical microscopy was used to track molecules and obtain brightness estimates that enabled calculation of crowding sensor copy numbers. The quantification of diffusing molecule trajectories paves the way for correlating subcellular processes and the physicochemical environment of cells under stress**.**

**Keywords:** molecular crowding, FRET sensor, fluorescence localization microscopy, osmotic stress, single-molecule, Slimfield

**Introduction**

Cellular response to external stress – nutrition deficiency, ionic strength, temperature, etc. – is regulated by signalling pathways that between them regulate almost all dynamic cell processes. Although many cellular stress responses have been uncovered which use movement and localization of key proteins for signal transduction and genomic control [1,2], the role of physical conditions within the cell, such as macromolecular crowding, is still unclear. Crowding – a local high density of biomolecules that can lead to a weak intermolecular attraction – has been shown to be a key parameter in the function and higher-order structures of proteins and nucleic acids [3] and protein/membrane interactions including aggregate formation [4] and it is therefore also of interest for stress response.

In recent years, Förster resonance energy transfer (FRET) based probes (consisting of a donor and acceptor fluorophore linked by an α-helix "spring") have been developed for use with *Saccharomyces cerevisiae* (budding yeast), the most common model organism for studying eukaryote stress response [5]. These probes may be sensitive to the presence of glucose and ATP [6], changes in crowding [7,8], ionic strength [9], or pH [10]. Commonly, these probes are genomically integrated [11], but it is possible also to nanofabricate a sensor and promote its uptake in cells, a process used to show the glucose starvation induced acidification of *S. ceverisiae* [12]. Crowding under dehydration has also been linked to liquid-liquid phase separation processes [13], and indeed in general the maintenance of crowding conditions inside the cell is a key parameter during the cell cycle and specifically in ageing [8].

Osmotic shock caused by a rapid change in extracellular ionic strength has a profound effect on molecular crowding within cells. Application of a high ionic strength medium overwhelms ion import and export and leads to a sudden decrease in cell volume with a concomitant increase in crowding [14]. In *S. cerevisiae* this leads to signaling slowdown [15,16] and regulation of the key osmoregulation protein Hog1 [17], a process which may be carbon-source dependent [18]. In *Escherichia coli,* it has been shown that, counterintuitively, an increase in crowding is associated with protein destabilization [19]. However, the extent to which subcellular crowding is heterogeneous is unclear, and it is unknown how or if crowding is clustered around subcellular features that use molecular machines or ion pumps to regulate local conditions.

Previous work has mostly used analysis techniques which rely on ensemble average fluorescence signal values over a cell or region of interest to calculate a ratiometric FRET value [7,8]. Rapid super-resolving single-molecule optical microscopy called Slimfield [20] however has been extensively used in single- and dual-color imaging experiments [1,21–23] to perform tracking of individual molecules and find their localization and estimate total copy numbers on a cell-by-cell basis [2]. Here, we measure the crowding in *S. cerevisiae* upon application of high and low ionic strength media with crGE [7], a crowding sensor which uses fluorescent protein mCerulean3 (a cyan fluorescent protein derivative) as a donor and mCitrine (yellow Fluorescent Protein derived) as the acceptor and which has been studied in budding yeast previously (Figure 1a) [8]. We use combined Slimfield and confocal microscopy to calculate approximate sensor copy numbers and demonstrate that the crowding within the cell is not dependent on local sensor concentration, and show crowding information in subcellular detail. Using realistic simulations of single-cell FRET heatmaps, we show that these experiments must be analyzed with care due to artifacts from regions with no FRET sensor present. We show these excluded volumes do not significantly affect measurement through whole-cell averages of the cytosolic content. Using whole-cell measurements, we show that glucose conditions affect crowding in basal and osmotic stress conditions, indicating a metabolic crowding response to glucose availability. Finally, we demonstrate the possibility of tracking individual FRET sensor molecules diffusing through a crowded cytosolic landscape.

**Abbreviations:**

WT: wild type; NaPi = Sodium phosphate buffer; FRET: Förster Resonance Energy Transfer; ConA: Concanavalin A; OD: optical density; YPD: Yeast Extract-Peptone-Dextrose medium; SD: synthetic defined medium; DIC: differential interference contrast

**Material and methods**

*Growth conditions and media*

Yeast cultures were grown in rich (YPD: 2% glucose, 2% peptone, 1% yeast extract) or synthetic (2% glucose, 1x yeast nitrogen base; including appropriate amino acids and bases) media (all percentages are w/v). The mCerulean3/mCitrine FRET crGE probe [7,8] under control of the *TEF1* promoter was sub-cloned into yeast integration vector pRS303, linearized with NsiI, then stably integrated into at the *HIS3* locus of wild-type BY4742 yeast by homologous recombination. Successful integrations were isolated on synthetic drop-out media (2% glucose, 1x yeast nitrogen base; 1x amino acid and base drop-out compositions lacking Histidine (SD –His, Formedium Ltd, UK). Single colony isolates were taken from solid agar following growth for 24-48 hours at 30°C and used to inoculate 8 mL of the appropriate media, then incubated at 30°C with agitation overnight. Cultures were grown to mid-log phase (OD$_{600}$ = 0.4 - 0.6) prior to harvesting 1 mL of culture, resuspending in 1 mL of imaging buffer (50 mM NaPi with or without 1 M NaCl) and preparation for optical microscopy.

*Sample preparation*

22x22 mm No. 1.5 BK7 glass coverslips (Menzel Glazer, Germany) were plasma cleaned in atmospheric plasma for 1 minute [24]. Two strips of double-sided tape were attached to a standard microscope slide approximately 5 mm apart to create a ~5-10 µL flow channel [25]. The coverslip was attached to the tape making a tunnel. The tunnels were washed with 20 µL of imaging buffer (50 mM NaPi) after which 20 µL of 1 mg/mL Concanavalin A (ConA) was introduced and allowed to incubate inverted for 5 minutes in a humidified chamber. The ConA was washed out with 200 µL of imaging buffer and 20 µL of cells were flowed in, and allowed to incubate for 5 minutes in the humidified chamber while inverted to promote adhesion to the coverslip. Finally, the channel was washed with 200 µL of imaging buffer and the tunnel sealed around the perimeter of the coverslip with nail varnish [26].

*Confocal microscopy and analysis*

Confocal microscopy was performed on a commercial laser scanning confocal microscope (LMS 710, Zeiss) attached to a commercial microscope body (Axio Imager2, Zeiss), using a Nikon Plan-Apochromat 63x NA 1.4 oil-immersion objective lens. Excitation and emission were measured with the following excitation lasers and imaging wavelength ranges: mCerulean 458nm and 454/515nm, mCitrine 514nm and 524/601 nm, FRET 458 nm and 458/601 nm with lasers set to 0.7% and 2.1% of maximum respectively. See Supplementary Table 1 for full imaging parameters.

Analysis of the confocal data was performed in Fiji (ImageJ 2.0.0-rc-69/1.52p/Java 1.8.0_172). Cell outlines were selected and segmented in the DIC channel either manually or using the Cell Magic wand plugin [27] and were saved as overlays in the original stack images. A bespoke ImageJ macro was used to extract the pixel intensities in each channel and calculate the ratiometric FRET measurement parameter *NFRET* [28] (see Supplementary Figure 1). The same macro was used to extract cell area. The intensities were corrected for autofluorescence by subtraction of the mean autofluorescence values found by imaging BY4742 wild type cells in experimental conditions. As in previous work [7] we defined the ratiometric FRET as $I_F/I_D$ where where *I$_F$* is the intensity in the FRET channel under donor excitation and *I$_D$* is the intensity in the donor channel under donor excitation. Using this approach, bleed-through corrections are not required since they are manifest as a constant offset in the ratiometric FRET distribution, while in line with previous work with this FRET sensor [7] cross-excitation can be neglected as it is a minimal contribution (around 4% of the excitation peak). For comparison, we also analysed our data with another common approach using normalized ratiometric FRET [8,28], defined as $NFRET = I_F/\sqrt{I_D I_A}$ where *I$_F$* and *I$_D$* are defined as above and *I$_A$* is the intensity in the acceptor channel under acceptor excitation. In general for *NFRET* analysis to be valid, intensities must be corrected for background, bleed-through, and cross-excitation of the acceptor under donor excitation. Here, we estimated bleed-through to be <7% in our confocal microscopy given the filter sets used. Comparing the uncorrected *NFRET* with the ratiometric FRET still indicated qualitative agreement without bleed-through correction. In all cases, we corrected for background noise which was accounted for by subtraction of the mean background of a region of interest in each confocal microscopy image, and autofluorescence which was taken to be the mean of a wild type dataset imaged under experimental conditions.

To generate heatmaps, the ratiometric FRET values in the stacks of images were calculated for each pixel rather than an average fluorescence intensity. Relative molecular stoichiometries of the dye were estimated either with conservation of energy considerations or by normalizing to the highest intensity pixel in the acceptor channel. For conservation of energy stoichiometry estimation, the conserved energy quantity was taken to be

$$E_{conserved} = hc \left( \frac{I_D}{\lambda_D} + \frac{I_F}{\lambda_A} + \frac{I_A}{\lambda_A} \right),$$

where $h$ is Planck's constant, $c$ is the speed of light, $I_D$, $I_A$, and $I_F$ are the intensities in the donor, acceptor, and FRET channels respectively, and $\lambda_D$ and $\lambda_A$ are the peak emission wavelengths of mCerulean3 and mCitrine respectively. The $E$ value was calculated for each pixel in the region of interest, and the resulting heatmap was normalized to the highest value to give an approximation of relative copy number.

*Cell fluorescence simulations*

Bespoke simulation software was written in Python 3 using SciPy [29], NumPy [30] and Matplotlib [31]. First, a canonical point spread function was found using the freely available Python library psf [32], and was downsampled to create a point spread function (PSF) cuboid unit of (xyz) 50x50x100 nm voxels. A randomly sized ellipsoidal cell [33] was placed in a 3D pixel grid where each pixel also had side length 50 nm, and within this was placed a spherical excluded volume to act as a vacuole. A point in this volume was selected and if found to be inside the cell but outside the excluded volume the locus was accepted. The canonical PSF intensity was then multiplied by a random factor between 0.9 and 1.1 to simulate variation in fluorophore brightness and a given fraction of the intensity was placed in the donor volume at the selected point, while the remaining intensity was given to the FRET volume. Finally, the canonical PSF intensity was multiplied by the ratio of the accepter to donor brightness, and again the brightness was scaled in the region 0.9-1.1. This process was performed until $10^5$ fluorophores had been accepted. Finally, every voxel was given a background noise value taken from a Gaussian distribution with parameters found experimentally, and within the cell the voxels were additionally given an autofluorescence value, again the Gaussian distribution of which was found from the confocal imaging data of wild type yeast in experimental conditions. To simulate the excluded membrane volume, the autofluorescence region was extended beyond the volume in which fluorophores were placed.

*Single-molecule microscopy and copy number estimation*

Single molecule microscopy was performed on a bespoke epifluorescence microscope described previously [1] modified to use 445 nm and 514 nm wavelength laser lines (Coherent Inc) at powers of around 13.5 mW and 14 mW respectively. Briefly, the microscope comprises a 100X oil-immersion Nikon Plan Apo 1.49 NA object lens, a CFP/YFP dichroic mirror (ZT442/514rpc, Chroma), and a 300 mm tube lens for a final resolution of 53 nm/pixel. Images were acquired using MicroManager using in-house developed control software [34], with the exposure time set to 10 ms using full EM gain. Since the 445 nm wavelength laser was continuous wave as opposed to digitally modulated, we began an acquisition of 100 frames and manually triggered the shutter which was programmed to open for 70 ms (7 acquisition frames), with the middle frame chosen for analysis to minimize shutter vibration effects. In the acceptor channel, 1000 frames were acquired again through manual shutter control. In all cases the donor/FRET channels were excited first followed by the acceptor channel. The center of the cell was focused using brightfield imaging and set for fluorescence acquisition. The blue and yellow channels were split and imaged side-by-side on a BSI Prime 95B. To analyze the Slimfield data, the first bright frames were used to avoid photobleaching effects. The cells were manually selected using ImageJ and analyzed with a custom Python 3.6 script. Results were plotted with Matplotlib. For heatmap generation, the brightfield frames were split vertically to create two separate images which underwent image registration using Scipy's Registration module. The calculated vector shift between channels was then applied to fluorescence images and heatmaps generated by calculating *NFRET* pixel-by-pixel. Thresholding was performed based on the $I_D$ and $I_A$ values as described subsequently. Single molecule fluorescence intensities and trajectories were determined using our home-written software (MATLAB, MathWorks) called ADEMSCode (version 2.8) [35]. Since single molecules were only visible in the acceptor channel, this channel was used to determine total visible copy number by dividing the integrated intensity within cells by the single molecule intensity. Confocal images for which the relative copy number had been determined were normalized by dividing by the integrated intensity [36,37] then were multiplied by the mean number of crGE sensors per cell [38] to give an estimate of absolute copy number at each pixel as has been done previously [36]. We note that for this analysis, neither ratiometric FRET nor *NFRET* was calculated, only the single-molecule intensity of the acceptor. Excitation of the donor at 514 nm wavelength is effectively zero compared to the detector noise so that cross-excitation and bleed-through in this case may be neglected, leaving only standard background correction necessary. All software used is openly available at https://sourceforge.net/projects/york-biophysics/files/Molecular%20crowding/

**Results and discussion**

**Measuring the increase in crowding and decrease in area due to osmotic shock**

We verified our strains and crowding sensor by initially quantifying molecular crowding and area changes in cells experiencing osmotic stress (Figure 1c and d). While vacuoles are clearly visible in fluorescence micrographs (see Figure 1 and Supplementary Figure 2) and expected to be an excluded volume [8], their presence in the whole-cell analysis does not significantly change the values or distribution measured and thus continuing analysis on a whole-cell level is justified (Figure 1b). At 0 M NaCl cells are in a low stress condition, but under exposure to 1 M NaCl crowding increases, evidenced by the ratiometric FRET increasing by 13.7% (Figure 1c) while measurable cell area is reduced by around 21% (Figure 1d) and thus volume by around half for a spherical cell, consistent with previously found values [14], due to water being mechanically forced from the cell by osmotic pressure [39].

**Glucose conditions during growth affects basal molecular crowding**

To investigate whether glucose availability during growth affects the molecular crowding at low osmotic stress, cells were grown at 1%, 2% or 4% glucose and imaged in 50 mM NaPi as described. Cells grown in 1% glucose showed the highest ratiometric FRET, with a small shift of -2.7% between

1% and 2% growth conditions (Figure 2a). When analyzing the same data with *NFRET* we find that the shift is approximately 0.3% if the comparison based on the mean *NFRET* value or of 0.6% if the median is used as the reference (Supplementary Figure 3a). The difference between 1% and 4% glucose conditions is more dramatic, with the ratiometric FRET reducing by 11.6% and the *NFRET* value reduced by 4.3% (Figure 2 and Supplementary Figure 3). Across conditions therefore we find a reduction in molecular crowding with increasing local glucose concentration. Although in the main text we include data from one set of experiments, we have taken multiple datasets and note that they all demonstrate a statistically significant difference in crowding between yeast grown in 1% and 4% glucose, though the difference between 1% and 2% is variable and would require further investigation (Supplementary Figure 4)..

Perturbation of the cells with 1 M NaCl also leads to different ratiometric FRET values. Specifically, the cells grown in 4% glucose undergo a higher relative shift in ratiometric FRET than those grown at 1% or 2% glucose, but the FRET ratio overall remains below that in the lower two glucose cases (Figure 2a). It appears that while the osmotic shock produces an increase in ratiometric FRET of 10-15% in each case, the underlying metabolic state of the yeast remains important in determining the final crowding state.

To cope with changes in external osmolarity, the high-osmolarity glycerol (HOG) pathway mediates regulation of glycerol production and water release, changing the properties of the cytoplasm and potentially impacting crowding [18,40]. Hyperosmotic conditions cause glycerol accumulation within the cell to maintain the cell size and water homeostasis [41]. Figure 1 illustrates a reduction of cell volume in response to rapid water removal due to sudden application of NaCl-induced hyper-osmotic shock. Such mechanical pressure appeared within the cell leads to an increase of the macromolecular crowding. However, the glucose conditions we used do not seem to have an influence on the cell volume (Figure 2b). We hypothesize that the overnight growth under lower glucose conditions (1% and 2%), leads to lower water content within the cell compared to that in cells grown in 4% glucose. As 4% glucose would mean higher external osmolarity than 1% and 2%, therefore, the changes in crowding may be caused not only by glycerol/water ratio shift but also trehalose and other osmolytes concentrations as has been suggested for respiring cells [18]. This is supported by the relative behavior of the different samples upon application of 1 M NaCl (Figure 2c). The FRET/mCerulean3 values increase by approximately the same proportion but remain lower in the 4% glucose case, indicating that there is a fundamental water:osmolyte ratio difference between the samples.

**The cell membrane is an excluded volume and leads to anomalous FRET values**

We began subcellular analysis of the ratiometric FRET data by generating a pixel-by-pixel estimate of the local FRET/mCerulean3 signal as described above. Interestingly, we noticed a ring of apparent high FRET/mCerulean3 values and thus apparently high crowding around the cell boundary, similar to that seen in the vacuole (Figure 3b), suggesting that the membrane is a similar excluded volume. To discern whether this was an effect due to autofluorescence and noise or due instead to a genuinely high crowding environment in the cell membrane, we simulated a yeast cell undergoing FRET as described in Materials and Methods.

With a straightforward simulation in the absence of an excluded volume in the membrane, we see no anomalous values as expected (Figure 3a). Upon addition of a simulated plasma membrane which emits fluorescence due to autofluorescence only, we recreate perfectly the confocal images, with a higher-FRET state apparently existing (Figure 3a). To remove this source of potential error in later tracking, we determined the ideal thresholding from autofluorescence data taken by fluorescence microscopy of wild type yeast. We thresholded the experimental data according to the donor channel fluorescence value $I_D$, comparing it to a test value $I_{D,T}$ where this value istaken to be the mean autofluorescence in the donor channel, plus a certain number of standard deviations. We note that for a Gaussian distribution 99.5% of the population is contained in the region $[\mu - 3\sigma, \mu + 3\sigma]$ and therefore do not go above mean plus or minus three standard deviations.

The effect of this thresholding in Figure 3b is striking – as the threshold increases the anomalous FRET/mCerulean3 values reduce until at mean plus three standard deviations they are effectively totally removed. This is clear evidence that the FRET probe has not penetrated the membrane and the higher values seen are coming from autofluorescence noise. Cross-checking this approach with simulated data (Figure 3c) we see almost identical behavior, and therefore determine that the membrane is indeed an excluded volume which should be rejected from single-molecule tracking analysis. Hereafter therefore all heatmaps are generated with the mean plus three sigma test, and pixels that fail to meet the criteria are set to 0.

Having accounted for the autofluorescence, the FRET/mCerulean3 values within each cell are relatively flat though there exists some apparent high values around the vacuolar region and the cell membrane for some of the cells in Figure 3, indicating that in the majority of the cytosol the yeast cells have assumed the physical equilibrium configuration of equally dispersed crowding but there may be some perturbations to this around membranes. A line profile of a single cell appears in Supplementary Figure 5 and shows a distribution around a mean though with some noise in the regions of the vacuole and membranes.

**Comparing relative stoichiometry with energy considerations and single-molecule copy number analysis**

Figure 4a and Supplementary Movie 1 shows the results of single molecule analysis of representative yeast acquisitions. The trajectories in Figure 4a were determined with bespoke single-molecule tracking software [35,42] adapted for two color imaging [43] are overlaid on the average cell intensity over the acquisition. Trajectories are uniformly short demonstrating either that the crGE acceptor fluorophores are prone to photoblinking, that they diffuse out of the field of view, or that trajectories collide and therefore are terminated by the software. Most likely it is a combination of the three possibilities. No tracks are seen entering the excluded volume vacuole region, as expected, and as there is a ring of fluorescence without tracks at the extreme edges of the cells it appears also that tracks do not enter the cell membrane as hypothesized in Figure 3. The *Isingle* value, the characteristic average brightness of a single dye molecule within the cellular environment [44] averaged across all spatial locations of cellular trajectories, was estimated and it was found that the mean copy number of crGE visible during Slimfield microscopy was approximately 99,000 molecules per cell, matching the number of fluorophores chosen for fluorescence simulations to mimic the appearance of the confocal images. This was done prior to single molecule analysis, and the closeness of the values indicates that this order of magnitude was reasonable. This analysis was possible with the acceptor channel detecting mCitrine signals, however, in the donor channel detecting mCerulean3 signals the single-molecule signal was below the background noise detection threshold. Even though the mCerulean3 fluorophore has improved photophysical properties compared to standard cyan fluorescent protein CFP, its absolute brightness is still lower than higher emission wavelength fluorescent proteins such as mCitrine [45].

Figure 4b and c shows the relative and absolute copy number estimation found by dividing the cell intensities by the total intensity of the field of view and multiplying by the number of cells and the mean number of crGE per cell. Both the acceptor intensity estimation and conservation of energy method give qualitatively similar results when comparing the relative stoichiometries, and when comparing the absolute values estimated they remain qualitatively very alike. The crGE sensor in all cases can be shown to be non-uniformly distributed inside the cell, while in Figure 3 we see that the FRET/mCerulean3 value is largely flat across the cell interior after correcting for autofluorescent noise. This leads to the conclusion that the ratiometric FRET values seen are not simply proportional to the local sensor copy number, as expected for a non-perturbative probe. Figure 4d shows the heatmap of this effect – within the cell it is clear that there are a range of FRET:copy number ratio values. It is unclear why the crGE copy number varies widely from cell to cell, but possible explanations for testing could be related to cell temporal or replicative age or stage in the replication cycle.

**Conclusions**

Probing the inner workings of cells requires an understanding not only of the biological molecules at work but the physical conditions in which they are found. Here, we used a genomically integrated FRET-based crowding sensor to assess molecular crowding as a function of osmotic stress, growth medium glucose concentration, and local sensor copy number. We found that a high glucose concentration (4%) during growth led to lower molecular crowding with no associated difference in cell area (Figure 2), pointing to a high-glucose metabolic state which fundamentally alters the physical conditions inside the cell, an effect strong enough to still be in evidence after the application of extreme osmotic stress in the form of 1 M NaCl. These effects were seen in both FRET/mCerulean3 analysis and *NFRET* calculation, despite a lack of bleed through correction, though the *NFRET* analysis appeared to be somewhat less sensitive, with a shift seen in crowding between 1% and 2% glucose growth conditions which was just above the significance level. We note that some Kernel Density Estimation (KDE) fits of these data look a little bimodal, possibly as a result of bet-hedging behavior as seen in budding yeast previously [46], where authors note that bet-hedging across the population is non-mutational and bimodal. In high-glucose conditions, sustained bimodal differences may help to guard the population against glucose deprivation. When analyzing confocal microscopy images on a pixel-by-pixel basis, we see that the crowding sensor is largely excluded from the cell membrane, and the autofluorescence in this region leads to apparently anomalous ratiometric FRET values as confirmed by bespoke fluorescence simulation (Figure 3). Cell membranes in the vicinity of cell walls are complex mesh-like environments as seen in other microbial organisms [47] that may necessitate the development of new biosensing probes with smaller dyes in order to probes these regions. Application of a simple threshold based on the wild type autofluorescence mean and standard deviation is sufficient to remove this artifact (Figure 3), although there is some evidence of higher crowding in these regions which are not artifact-derived. With physical considerations such as conservation of energy, it is possible to map the local distribution of fluorophores relative to a high intensity area of the cell. Both methods used here give qualitatively very similar results, but further *in vitro* work would be advisable to rigorously and quantitatively assess the approaches. The similarities seen here would suggest that both techniques are likely to have similar accuracy. The ratio of FRET/mCerulean3 to relative copy numbers demonstrate that the two are not simply proportional and the sensor is therefore measuring the crowding of the cell and not the crowding due to the sensor. We also note that this indicates that there is little intermolecular FRET under these conditions, which would confound analysis. Mean separation of the sensors in these conditions is on the order of tens of nm, a distance too great for effective FRET which supports this conclusion further.

Slimfield microscopy was used to image single fluorophores diffusing through the cell, and from these trajectories characteristic *Isingle* values of the acceptor fluorophore were obtained, leading to an estimation of absolute copy numbers within the cell on a cell-by-cell basis, with a mean copy number estimate of ~$10^5$ molecules per cell field of view. This value is an approximate one and the total copy number estimate will be affected by various experimental and physiological factors, for example degradation kinetics, yeast background, and copy number estimation method (Slimfield microscopy *vs.* immunoblotting a known standard, for example). Care should therefore be taken when comparing this abundance to other work. Short traces have been obtained but the overabundance of fluorophores alongside photoblinking and z-axis diffusion limit the trajectory length (Figure 4). In future work, the crGE sensor could be expressed under control of a suite of inducible and constitutive promoters to modulate the number of fluorophores. This should allow tracking of longer trajectories and a single fluorophore's path through the crowding landscape could then be found. Ultimately, cluster analysis and single-molecule tracking will be a powerful tool to relate the physical conditions within the cell with biological processes such as liquid-liquid phase separation, protein aggregation, and cell division.

**Funding sources**

This project has received funding from the European Union's Horizon 2020 research and innovation programme under the Marie Skłodowska Curie grant agreement no. 764591 (SynCrop), the Leverhulme Trust (reference RPG-2019-156), and BBSRC (reference BB/R001235/1), Royal Society Newton International Fellowship Alumni (AL/191025), and Wellcome Trust and the Royal Society grant no. 204636/Z/16/Z.

**Acknowledgements:**

We thank Arnold J Boersma (DWI-Leibniz Institute for Interactive Materials, Aachen, Germany) and Bert Poolman (University of Groningen, The Netherlands) for plasmid donation, Dr Ji-Eun Lee for assistance with Slimfield microscopy, and Karen Hogg, Grant Calder, Graeme Park, and Karen Hodgkinson (Bioscience Technology Facility, University of York) for support with confocal microscopy.

**Figures with captions**

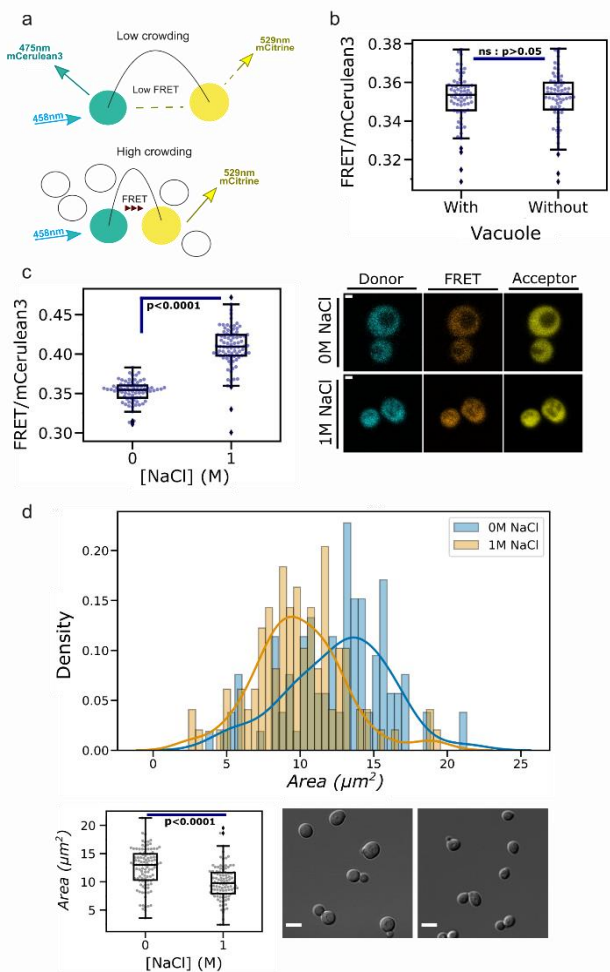

**Figure 1**

a) Schematic representation of the crGE dynamic FRET sensor. In low crowding, the donor and acceptor are separated such that non-radiative energy transfer is low, energy is emitted by the donor fluorophore, and low FRET efficiency is observed. A crowded environment will bring the two dyes closer to each other promoting non-radiative energy transfer, emission by the acceptor fluorophore, and hence a higher FRET efficiency.

b) Jitter plot of ratiometric FRET values for *S. cerevisiae* grown in 2% glucose and imaged with 0 M NaCl with and without including the vacuole in analysis. Boxes represent the interquartile range (IQR), with bars extending 1.5*IQR from the upper and lower quartile. Diamonds indicate data outside this range. Student's *t*-test shown with p=0.05.

c) Left: jitter plot of FRET/mCerulean3 values for yeast grown in 2% glucose and measured in no (0 M) and high (1 M) NaCl conditions. Box plot and significance testing as in panel b. Right: fluorescence micrographs of crGE in both 0 M and 1 M NaCl with vacuoles visible. Scale bar: 1 μm.

d) Histogram of cell area imaged in 0 M and 1 M conditions and grown in 2% glucose. Area is given in μm², with data fit by Kernel Density Estimation (KDE) [48] using the seaborn library function *distplot*. KDE bandwidth is set using the rule-of-thumb Scott's Rule. Below left: jitter plot of the same data with box plotting and significance testing as in panel b. Right: DIC images of the cells in low (0 M) and high (1 M) ionic strength buffer. Scale bars: 5 μm.

**Figure 2**:

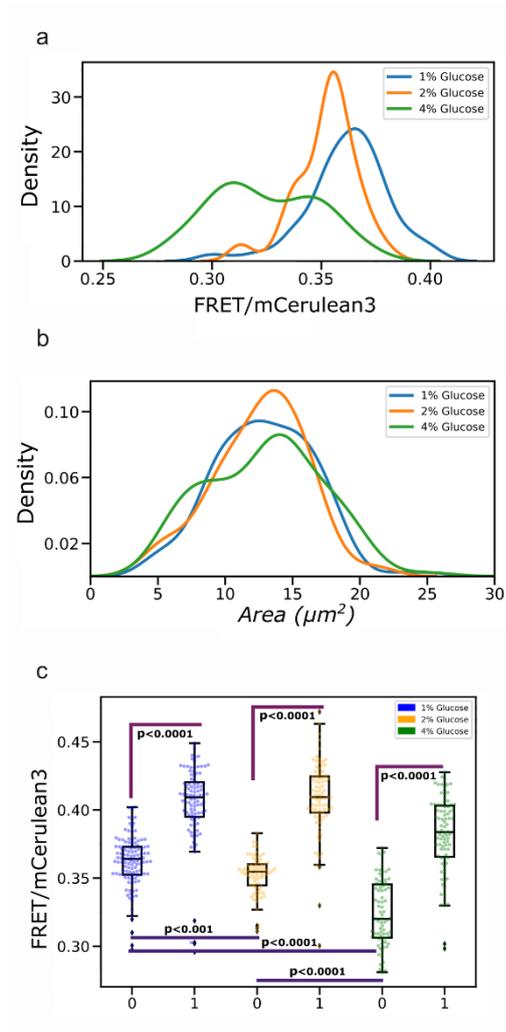

a) Kernel density estimates (KDEs) of the ratiometric FRET distribution for cells grown in 1%, 2%, and 4% glucose conditions and imaged at 0 M NaCl. Data is fit as in Figure 1d. For all conditions N>100.

b) KDE of the cell size distribution, with area in µm$^2$, for cells grown in 1%, 2%, and 4% glucose and imaged at 0 M NaCl. Lines are given by a KDE fit.

c) Jitter plot showing FRET/mCerulean3 at high (1 M) and low (0 M) salt concentration for cells grown at 1%, 2%, and 4% glucose and imaged in 50 mM NaPi. Here as we perform 6 *t*-tests we corrected the accepted p value via the Bonferroni method [49], thus here p=0.0083. Boxes represent the interquartile range (IQR), with bars extending 1.5*IQR from the upper and lower quartile. Diamonds indicate data outside this range.

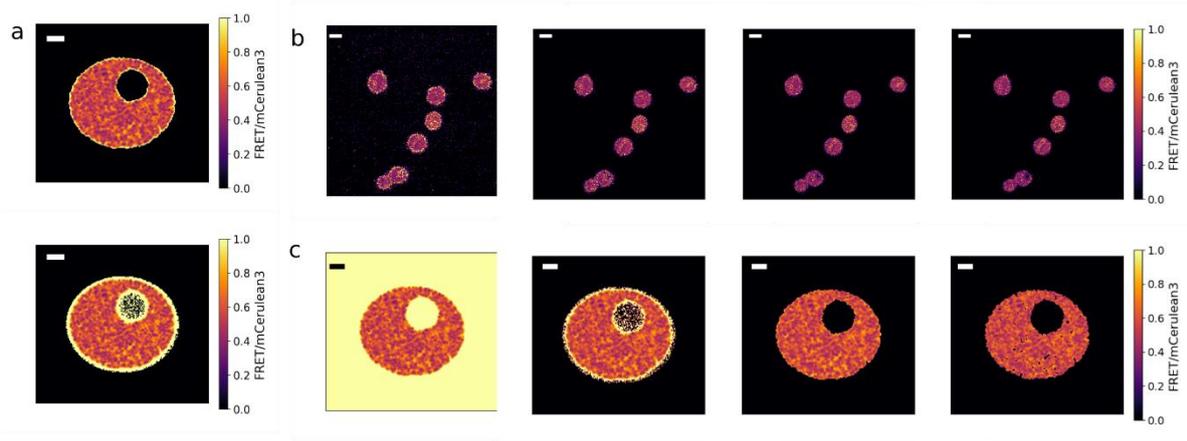

**Figure 3:**

a) Simulating a yeast cell without and with an outer membrane acting as an excluded volume (upper and lower panels respectively). Addition of the excluded volumes leads to a characteristic high-FRET ring, an anomaly caused by different autofluorescence values in each channel. Scale bars: 1 µm

b) Confocal data of crGE in 0 M NaCl and analyzed pixel-by-pixel shows a high-FRET ring at the cell edge. This may be segmented out by only including pixels where $I_D > I_{D,T}$ where $I_{D,T}$ are the mean donor autofluorescence values plus a number of standard deviations in the pixel autofluorescence values. Going left to right, these are thresholding with the mean only, thresholding with the means plus one standard deviation, thresholding with means plus two standard deviations, and thresholding with means plus three standard deviations. We see that thresholding in the final case removes >99.5% of autofluorescence-only pixels and effectively removes the anomalous ring. Scale bars: 5 µm.

c) Results of thresholding simulated data as described for panel b. In the final panel with mean plus three standard deviations the simulated cell shows some over thresholding with genuine pixel excluded, while the mean plus two standard deviations shows almost identical characteristics to the cell simulated without an excluded membrane volume in panel a. Scale bars: 1 µm

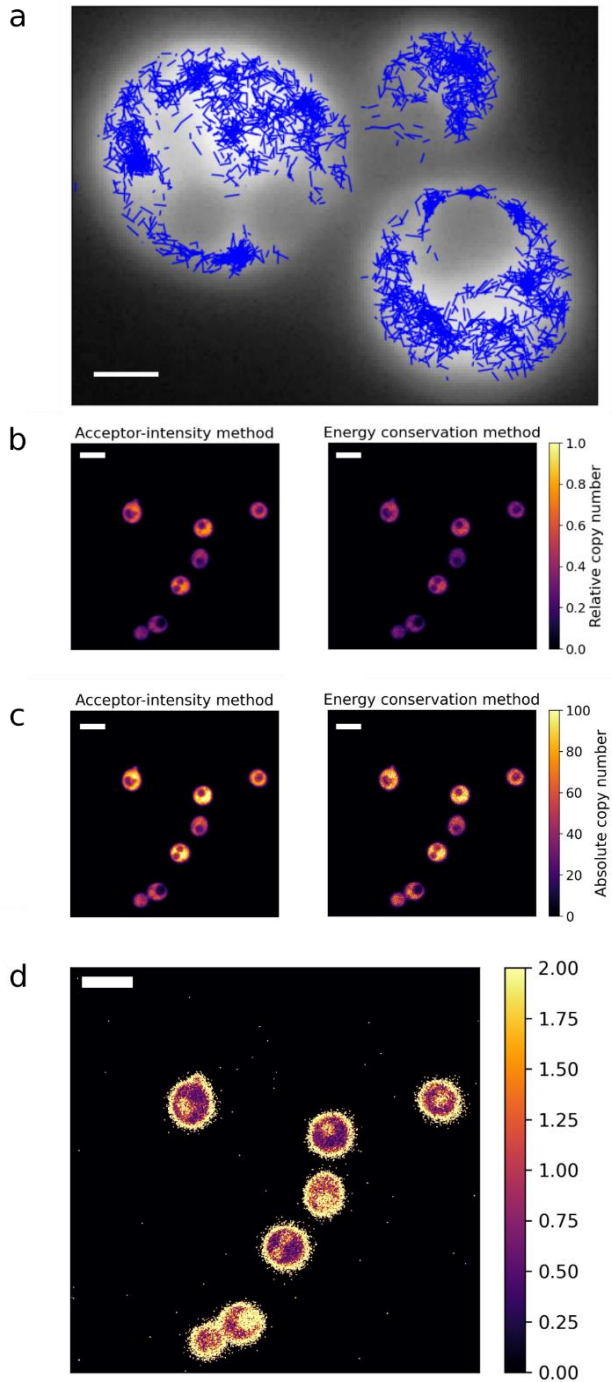

**Figure 4:**

a) Single-molecule trajectories (blue) identified by ADEMSCode, overlaid on frame average Slimfield fluorescence images (gray). Trajectories are uniformly short due to the number of molecules identified and their trajectories colliding. As expected, no tracks enter the excluded vacuole region. Scale bar: 1 µm.

b) Estimates of relative copy number found through the acceptor intensity method (left) and conservation of energy (right). Scale bars: 5 µm.

c) Estimated absolute copy number found by assuming each cell has the mean number of fluorophores present. Left: acceptor-intensity method; right: conservation of energy. Scale bars: 5 µm.

d) Heatmap of the ratio given by dividing the ratiometric FRET by the relative copy number found in panel b. Scale bar: 5 µm.

# Molecular crowding in single eukaryotic cells: using cell environment biosensing and single-molecule optical microscopy to probe dependence on extracellular ionic strength, local glucose conditions, and sensor copy number

Jack W Shepherd[1,a,b], Sarah Lecinski[1,a], Jasmine Wragg[c], Sviatlana Shashkova[a,d], Chris MacDonald[b], Mark C Leake[a,b]

[1]These authors contributed equally

[a] Department of Physics, University of York, York, YO10 5DD

[b] Department of Biology, University of York, York, YO10 5DD

[c] School of Natural Sciences, University of York, York, YO10 5DD

[d] Department of Microbiology and Immunology, Institute for Biomedicine, Sahlgrenska Academy, University of Gothenburg, 405 30 Gothenburg, Sweden

For correspondence: Jack W Shepherd, Department of Physics, University of York, York, YO10 5DD. E-mail: jack.shepherd@york.ac.uk

| Image size | 512x512 (pixels) // 41.52x41.52(um) | |
|---|---|---|
| Bit dept | 16 bit | |
| Lasers | 458nm (2.1%, 0.23uW) and 514nm (0.7%, 0.23uW) | |
| Pinhole | 0.86 Airy units, 0.7um section | |
| Scan direction | Unidirectional | |
| Detector digital gain | 1 | |
| mCerulean3 Excitation/Emission | 458nm/485nm | Detection 454-515 |
| Fret Ex/Em | 458nm | Detection 524-601 |
| mCitrine Ex/Em | 514nm/563nm | Detection 524-601 |
| Numerical aperture | 1.4 | |
| Scan speed (per frame) | 1.54s | |

**Supplementary Table 1:** Confocal microscopy imaging parameters.

a

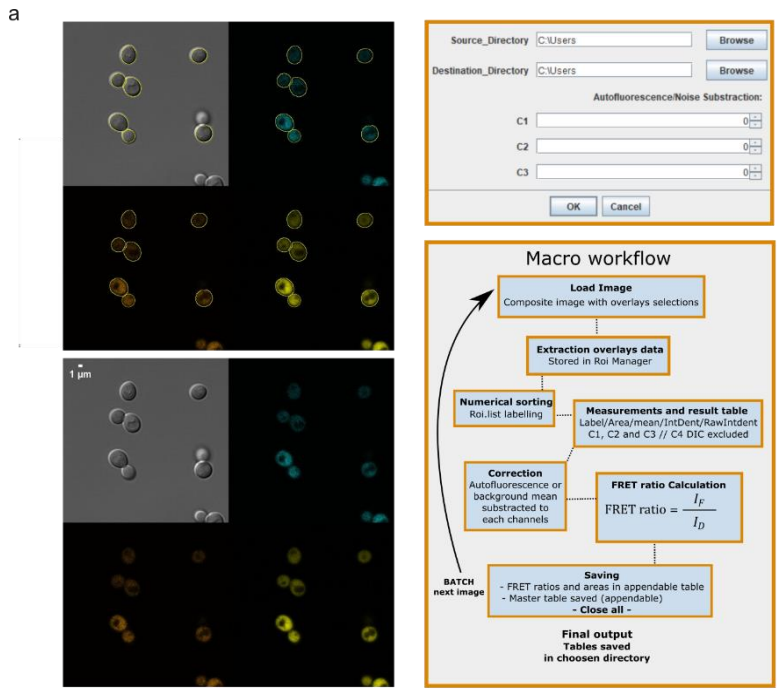

b

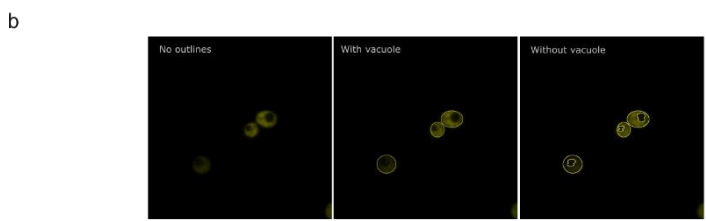

| Vacuole | With | Without |
|---|---|---|
| Median | 0.3536 | 0.3540 |
| Mean | 0.3515 | 0.3517 |
| T-test pValue | 9.38E-01 | |

**Supplementary Figure 1: GUI and workflow description of the ImageJ macro used to analyze data from the confocal microscope. Yeast segmentation and vacuole exclusion.**

a) Visualisation composite image aquired, cells outlines generated either with selection tool or the cell magic wand tool plugin and saved as metadata in the form of overlays (inactive selections), the image is composed of four channels, C1(blue): mCerulean3 exitation, the donor; C2(orange): the FRET channel; C3(yellow): mCitrine, the acceptor. Next to the images, the workflow and Graphical User Interface (GUI) of the diolague box generated when the macro is run.

b) Yeast cells segmentation and removal of the vacuole area: composite selection to exclude the vacuole generated in Fiji (Imagej). The table shows statistical properties of the two distributions including the p value taken from Student's *t*-test.

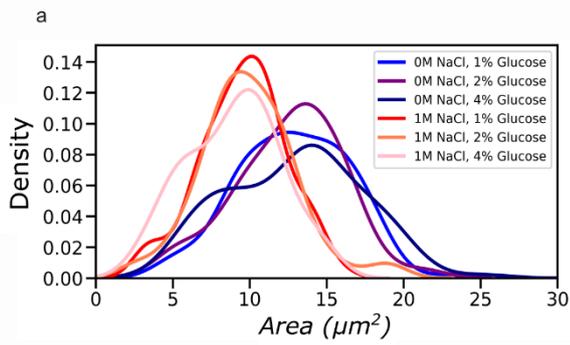

| 0M NaCl | CrGE1 | CrGE2 | CrGE4 |
|---|---|---|---|
| Median area | 12.579 | 13.023 | 13.555 |
| Mean area | 12.898 | 12.559 | 12.921 |
| pValue : paired T-test between CrGE1 and CrGE2 | 0.457 | | |
| pValue : paired T-test between CrGE2 and CrGE4 | | 0.5035 | |
| pValue : paired T-test between CrGE1 and CrGE4 | 0.9634 | | |

| 1M NaCl | CrGE1 | CrGE2 | CrGE4 |
|---|---|---|---|
| Median area | 9.76 | 9.787 | 9.267 |
| Mean area | 9.567 | 9.896 | 8.918 |
| pValue : paired T-test between CrGE1 and CrGE2 | 0.4143 | | |
| pValue : paired T-test between CrGE2 and CrGE4 | | 0.0295 | |
| pValue : paired T-test between CrGE1 and CrGE4 | 0.1192 | | |

**Osmotic shock FRET ratio & area**

| 0M to 1M NaCl | CrGE1 | CrGE2 | CrGE4 |
|---|---|---|---|
| Mean Area | 25.00% | -21.20% | -30.90% |
| T-test pValue | 5.47E-14 | 1.86E-08 | 9.52E-12 |

| 0M to 1M NaCl | CrGE1 | CrGE2 | CrGE4 |
|---|---|---|---|
| Mean shift FRET/mCerulean3 (%) | 10.55 | 13.76 | 15.07 |
| T-test pValue | 1.43E-29 | 1.69E-35 | 3.61E-30 |

**Basal condition FRET ratio**

| 0M NaCl | CrGE1 | CrGE2 | CrGE4 |
|---|---|---|---|
| Median ratio FRET/mCerulean3 | 0.3640 | 0.3548 | 0.3201 |
| Mean ratio FRET/mCerulean3 | 0.3621 | 0.3524 | 0.3243 |
| Median Shift (%) | 0.00 | -2.61 | -13.71 |
| Mean Shift (%) | 0.00 | -2.76 | -11.67 |
| pValue : paired T-test between CrGE1 and CrGE2 | 0.000168 | | |
| pValue : paired T-test between CrGE2 and CrGE4 | | 3.45E-15 | |
| pValue : paired T-test between CrGE1 and CrGE4 | 5.07E-26 | | |

**Supplementary Figure 2 :**

Area measurements at 0 M and 1 M NaCl for all glucose conditions (1%, 2% and 4%). Lines are KDE fits. Below, statistical tests between conditions using Student's *t*-test.

**Supplementary Figure 3**

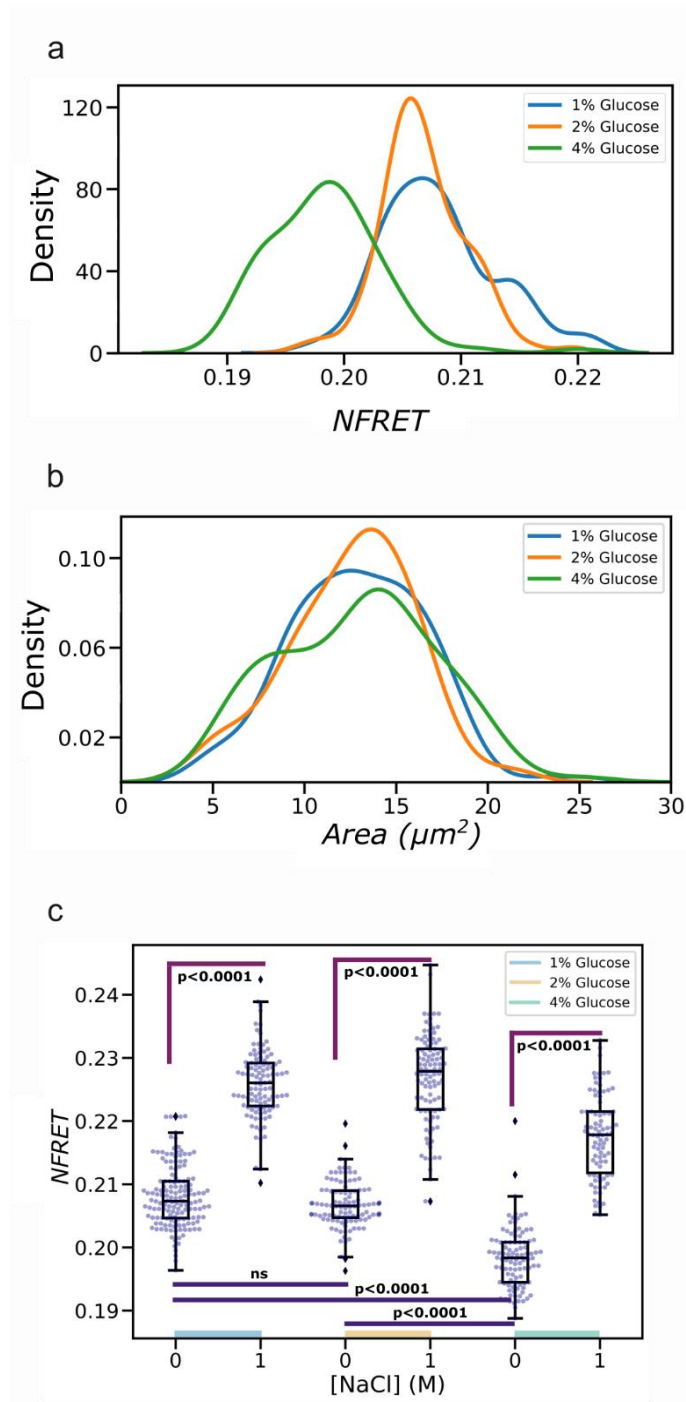

a) Kernel density estimates (KDEs) of the *NFRET* distribution for cells grown in 1%, 2%, and 4% glucose conditions and imaged at 0 M NaCl. Data is fit as in Figure 1d. For all conditions N>100.

b) KDE of the cell size distribution, with area in µm$^2$, for cells grown in 1%, 2%, and 4% glucose and imaged at 0 M NaCl. Lines are given by a KDE fit.

c) Jitter plot showing *NFRET* at high (1 M) and low (0 M) salt concentration for cells grown at 1%, 2%, and 4% glucose and imaged in 50 mM NaPi. Here as in Figure 2 we have corrected the p-value with Bonferroni's method.

**Supplementary Figure 4**

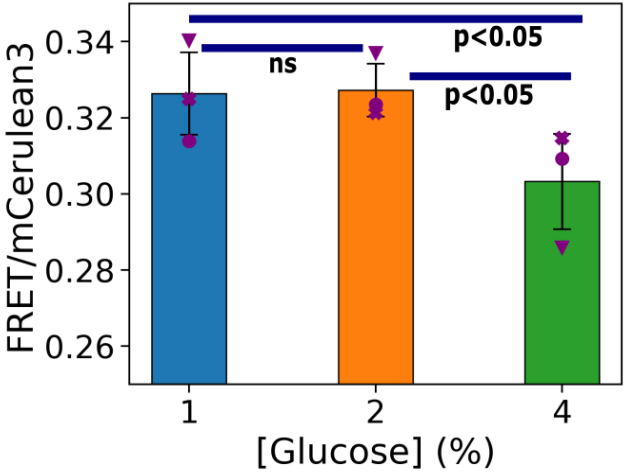

Ratiometric FRET dependence on glucose as measured during 3 repeats of the experiment. Circles, triangles and crosses indicate the experiment was performed on the same day with the same overnight liquid culture.

**Supplementary Figure 5**

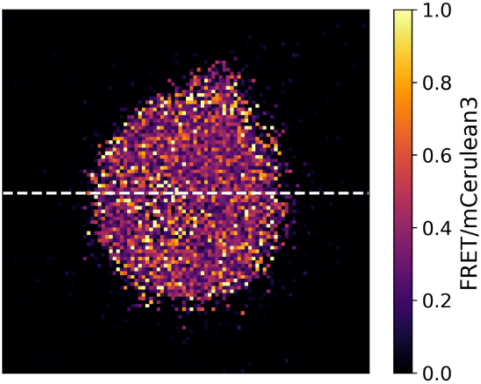

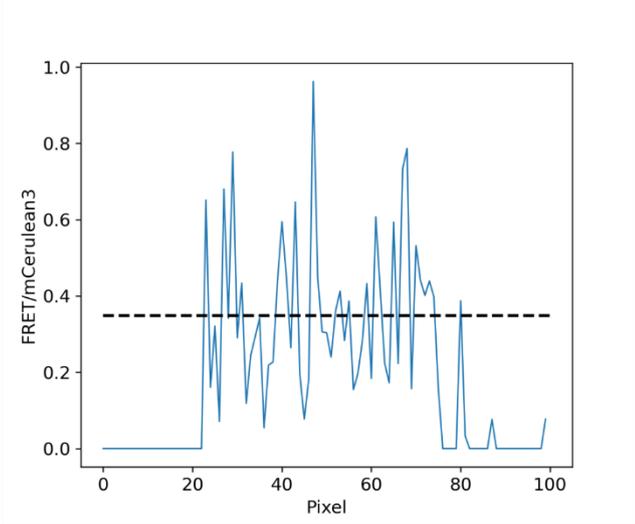

Top: A single cell isolated from the FRET/mCerulean3 heatmap in Figure 3. The dashed white line indicates the line profile taken. Scale bar: 1 µm. Below: ratiometric FRET line profile.

**Supplementary Movies**

**Supplementary Movie 1**

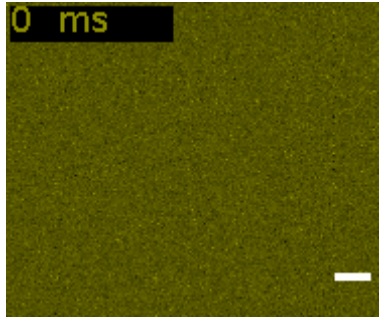

Movie showing diffusion of individual mCitrine fluorophores when under excitation with the 514 nm wavelength laser, 5ms per frame. Bar: 1 μm.